\documentclass[useAMS,usenatbib]{mn2e}
\usepackage{graphicx,fleqn,times,color}
\usepackage{lscape}

\def\today{\ifcase\month\or
 January\or February\or March\or April\or May\or June\or
 July\or August\or September\or October\or November\or
 December\fi\space\number\day, \number\year}

\def\todmy{\number\day\space\ifcase\month\or
 January\or February\or March\or April\or May\or June\or
 July\or August\or September\or October\or November\or
 December\fi\space\number\year}

\title[Metallicity-dependent kinematics and morphology of the
  Milky Way bulge]{Metallicity-dependendent kinematics and morphology of the Milky Way bulge}

\author[E.~Athanassoula, S. Rodionov, N. Prantzos]
{ E.~Athanassoula\thanks{E-mail: lia@lam.fr}$^1$,
 S. A.~Rodionov$^1$, N. Prantzos$^2$\\ 
\\
1 Aix Marseille Univ, CNRS, LAM, Laboratoire d'Astrophysique de Marseille, Marseille,
France\\
2 Institut d'Astrophysique de Paris (IAP), UMR7095 CNRS, Univ. Pierre \&
Marie Curie, 98bis Bd. Arago, 75014 Paris, France\\
}

\begin{document}

\date{Accepted . Received -}

\pagerange{\pageref{firstpage}--\pageref{lastpage}} \pubyear{2012}

\maketitle

\label{firstpage} 
\begin{abstract}
We use N-body chemo-dynamic simulations to study the coupling between
morphology, kinematics and metallicity of 
the bar/bulge region of our Galaxy. We make qualitative 
comparisons of our results with available observations and find very good
agreement. We conclude that this region is complex, since
it comprises several stellar components with different properties --
i.e. a boxy/peanut bulge, thin and thick 
disc components, and, to lesser extents, a disky pseudobulge, a stellar
halo and a small classical bulge -- all cohabiting in
dynamical equilibrium. Our models show strong links
between kinematics and metallicity, or morphology and metallicity,
as already suggested by a number of recent observations. We discuss
and explain these links.
\end{abstract}

\begin{keywords}
Galaxy: bulge -- Galaxy: stellar content -- Galaxy: structure -- Galaxy:
kinematics and dynamics -- Galaxy: evolution
\end{keywords}

\section{Introduction}
\indent

Observations have recently revealed an interesting coupling
between kinematics and metallicity of stars in the Milky Way (MW) bar/bulge
region.
\cite{Babusiaux.16} (her Fig. 4, see also \citealt{Babusiaux.P.10}),
reviewing the link between metallicity and kinematics, collected
data from a number of sources 
and plotted the velocity dispersion ($\sigma$) versus the absolute value of the
latitude ($b$). She found that for low metallicity stars, $\sigma$ shows
very little, if any, trend with $|b|$, while for high metallicity stars
$\sigma$  clearly decreases with increasing $|b|$. 

\citealt{Ness.P.13b} 
(hereafter N13b; see the lower panels of their figure 6) binned the ARGOS data by
metallicity, and plotted the 
velocity dispersion as a function of longitude $l$. This revealed that 
the higher
metallicity stars (0$<$[Fe/H]) have two clear trends: First, stars at
$l$=0$^{\circ}$ have high velocity dispersions which decrease with
increasing $|l|$, and, second, stars  at low latitude
($|b|$=5$^{\circ}$) have a larger velocity dispersion than stars at
high latitude ($|b|$=10$^{\circ}$). They also showed that the     
lower metallicity stars (-1.0$<$[Fe/H]$<$-0.5) have higher
velocity dispersions than the higher metallicity stars and also  
that the velocity dispersion depends only little on
longitude or latitude.

Just after this letter was first
  submitted, \citeauthor{Zasowski.NGMJM.16}
  (\citeyear{Zasowski.NGMJM.16}, hereafter Z16) presented APOGEE data
  for the MW  
  bulge. Given the relevance of these data to our results we added a 
  posteriori a short discussion of them.

We present here a theoretical study of this interesting chemo-kinematic
coupling. Previous simulation studies of metallicity in the bar/bulge region
\citep{Bekki.Tsujimoto.11, Martinez.VG.13, diMatteo.P.15} used pure
N-body simulations and thus included neither gas nor star
formation. Nevertheless, by assuming an initial metallicity radial distribution,
they were able to study its redistribution due to bar formation and
evolution. 
Here, we avoid  
such a short-cut, and use
a coupled chemical-kinematical-morphological approach, based on an N-body
simulation obtained with a code including both gas and star formation, 
coupled
to a chemical evolution code which follows the
distribution of the chemical elements as a function of time and  
location in the galaxy.  
The results of this letter were presented in two
meetings on our Galaxy, one in Paris (19 - 23/09/2016)\footnote 
{http://www.iap.fr/vie\_scientifique/ateliers/MilkyWay\_Workshop/2016/
and then click on program} 
 and the other in
Canberra (21 - 25/11/2016)\footnote{https://www.aao.gov.au/files/conferences/Lia\_GASP16\_0.pdf}. 


\section{Simulations}
\label{sec:sims}

\subsection{General context}
\label{subsec:general}

Our disc galaxy formation model has been described in detail by
\citeauthor{Athanassoula.RPL.16} (\citeyear{Athanassoula.RPL.16},
  hereafter A16). We will present it here only very briefly. We assume
  that the galaxy we model underwent a major merger about 8 -- 10 Gyr
  ago and start our simulations with two spherical protogalaxies
composed solely of dark matter and gas, while stars form all through
the simulation. These two protogalaxies are set on an orbit bringing
them to a collision. The stars born before the merging
undergo violent relaxation and form a spheroidal, centrally concentrated  
object. The stars born during the merging are strongly shuffled by the
quickly varying potential and form mainly a thick disc or an extended
stellar halo. Up to this point the evolution is merger driven. After the
merger settles down 
the evolution becomes secular, and the thin disc starts forming 
from the gas accreted from
the gaseous halo. Some of it may thicken and contribute to the thick
disc, but, as gas accretion continues all through the simulation, a thin
disc of relatively younger stars is always present.    

\subsection{Code}
\label{subsec:code}

The code describing the dynamical evolution is based on \textsc{gadget3}
(a Tree SPH code, \citealt{Springel.Hernquist.02, Springel.Hernquist.03})
and is described in A16 and in Rodionov et al. (subm.). The mass of
the baryonic particles is $10^4 M_\odot$, and that of the
dark matter ones $4 \times 10^4 M_\odot$, with 
10 and 17.5 million particles in each of these components,
respectively. The softening is 25 pc.

For the chemical part we adopt the Single Stellar Population (SSP)
formalism, where a single simulation particle represents a 
population of stars of the same age. Within that ``stellar particle",
stars are distributed according to the IMF of \cite{Kroupa.02}. They
release their ejecta after a finite, mass dependent, lifetime. Yields
for 12 selected elements are metallicity dependent and are taken from 
\cite{Nomoto.KT.13} 
for single stars and from \cite{Iwamoto.p.99} for SNIa
(see Appendix C in \citealt{Kubryk.PA.15} for details). A more detailed
description 
will be given elsewhere (Rodionov et al., in prep.).

There is one crucial free parameter in the present application of our
code, namely the metallicity at the
beginning of our simulation. If we had a cosmological simulation we
would have started with zero metals. However, in order to fully follow the
dynamics and evolution of the bar/bulge region we need to have a
dynamical simulation which does not start at the time of the Big Bang
and the formation of the
large scale structure. It just starts from the
progenitors of the last major merger and we need to assume that some
of the elements are 
already formed. In this letter we use 
an initial
metallicity of [Fe/H]=-1 \citep{Wolfe.GP.05}. More on
different choices will be included in 
Rodionov et al. (in prep.).

\subsection{Model}
\label{subsec:model}

Simulation mdf732, described in some detail in A16, has
properties that make it a reasonable choice for a qualitative model of the
bar/bulge region of our Galaxy. In particular it has a 
classical bulge with only 9 -- 12\% of the total stellar mass and a
bar of roughly the correct size, with a boxy/peanut inner part. We thus
reran it using 27.5 million particles to enhance the signal-to-noise,
because we need to split the data by metallicity, $b$ and $l$. 
We use here the snapshot of this new simulation at time $t$=10 Gyr, but
we made sure that  
other late times in the simulation around this one gave very similar results.

As described in A16, there are two times which can be considered as
landmark times for the disc galaxy formation. The first is the
beginning of the merging period, or merging
time, $t_{bm}$, which is the time beyond which the distance between
the centres of the two protogalaxies becomes and stays less than 1
kpc (A16). The second time is $t_{bd}$, the time at which the thin
disc starts forming. The values of these times for our simulations are
1.4 and 2.2 Gyr, respectively (A16).  

Although we do not claim to have a full model of our Galaxy, our model
is a
sufficiently reasonable approximation to use for the
present qualitative comparison. 
Here we follow
\citeauthor{Ness.P.13a} and consider only stars within a cylindrical
radius of 3.5 kpc from the centre, 
so as to concentrate on the bar/bulge region. 
Except for Fig. ~\ref{fig:sideon4} and its discussion, we use
everywhere the geometry of our
Galaxy, placing the bar major axis at 27$^{\circ}$ to the line from the
Sun to the Galactic centre and assuming that the Sun is on the
equatorial plane at a distance of 8 kpc from the centre.

\section{Results}
\label{sec:results}

In all the following analysis we use three metallicity bins, namely low
metallicity (-1.0$<$[Fe/H]$<$-0.5, hereafter LM), intermediate
(-0.5$<$[Fe/H]$<$0, hereafter IM) 
and high metallicity (0$<$[Fe/H]$<$0.5, hereafter HM)
bins, with one exception, namely in Sect.~\ref{subsubsec:ARGOS-kinem},
where the high metallicity bin is defined as 0$<$[Fe/H], to follow
N13b. Furthermore, in Sects.~\ref{subsec:age-jj-plot} and
\ref{subsec:spatial-distrib} we add a yet
higher metallicity bin with  0.5$<$[Fe/H]$<$1 (hereafter HHM).

\subsection{Metallicity dependent kinematics}
\label{subsec:kinem_orb}
 
\subsubsection{Comparison to the Babusiaux plot}
\label{subsubsec:babplot}
 
\begin{figure*}
  \centering
  \includegraphics[scale=0.26]{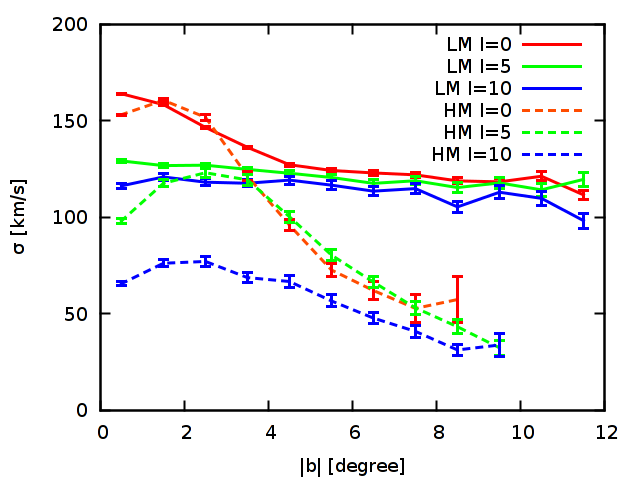}
  \includegraphics[scale=0.26]{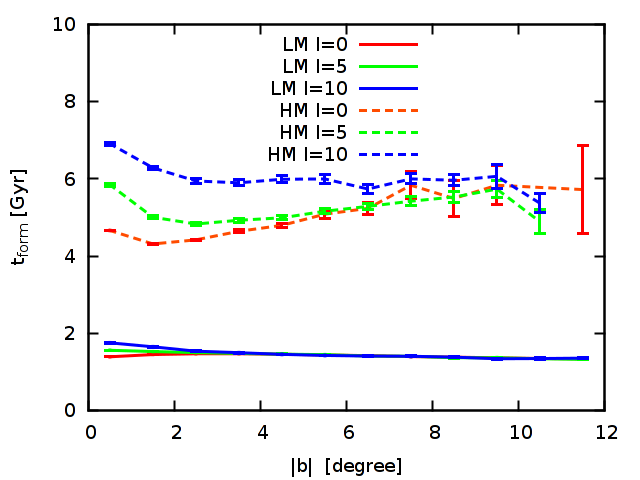}
  \includegraphics[scale=0.34]{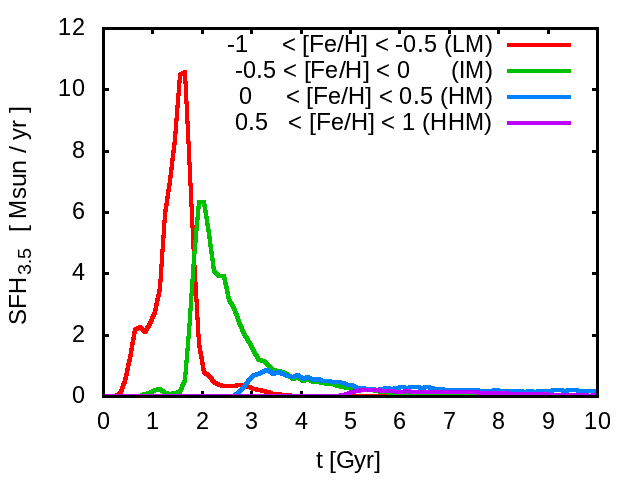}
  \caption{Left: Velocity dispersion as a function of galactic longitude for
    two metallicity bins, one with LM stars 
    and the other with HM stars. For each metallicity bin we average
    values from positive and negative $b$ and we show
    longitudes with 
    $l$=0$^{\circ}$, 5$^{\circ}$, and 10$^{\circ}$.  
    Middle: Mean star formation time as a function of galactic
    latitude. The layout, as 
    well as the split in two metallicity groups is the same as for
    the left panel. 
    Right: Star Formation History within the considered region ($R$$<$3.5
    kpc) as a function of time, separately for  
    stars in LM, IM, HM and HHM metallicity bins.
    }
  \label{fig:various}
\end{figure*}

To show the link between metallicity and kinematics, 
\cite{Babusiaux.16} plotted the dispersion of the line-of-sight
velocity as a function of latitude and constrains her plot to
stars around $l$=0$^{\circ}$ and from either the LM, or the
HM metallicity ranges, i.e. neglecting all stars with IM
metallicities. 
We repeated this for our simulation and three 
values of the longitude and give the results in the left panel of
Fig.~\ref{fig:various}. We find that,
for LM stars, $\sigma$ takes higher values than for HM ones.
Furthermore, for LM stars $\sigma$ varies little with $|b|$ or
$|l|$, while for HM stars it decreases considerably with increasing 
$|b|$. Note also that this decrease is strongest for $l$=0$^{\circ}$,
and weakens with increasing $|l|$. We thus conclude that
there is an excellent qualitative 
agreement between our simulations and observations (and even a
reasonable quantitative one, as can be seen by comparing our
figure with figure 4 of \citealt{Babusiaux.16}).

\subsubsection{Comparison to ARGOS and APOGEE results}
\label{subsubsec:ARGOS-kinem}
 
\begin{figure*}
  \centering
  \includegraphics[scale=0.38]{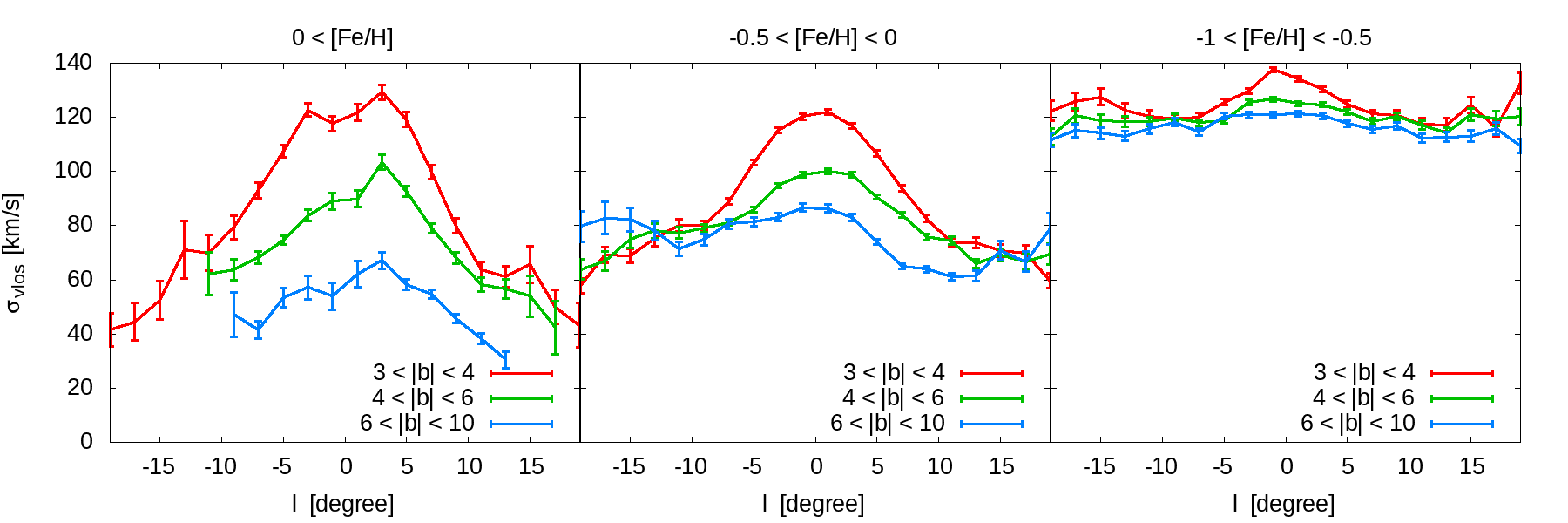}
  \caption{Velocity dispersion as a function of Galactic longitude for
    three metallicity bins. 
    }
  \label{fig:argos-plot}
\end{figure*}

Fig.~\ref{fig:argos-plot} shows the velocity dispersion as a function
of $l$ for 
three metallicity bins, i.e. from left to right, stars with
0$<$[Fe/H], IM and LM stars 
respectively, 
i.e. the same bins as in N13b, so that it can be
directly compared to the lower panels of figure 6 of that paper.
LM stars have in general higher values of $\sigma$ 
than HM ones.  
For HM stars, $\sigma$ is much larger at
small $|b|$ than at large $|b|$. This difference with $|b|$ is
very strong for l=0$^{\circ}$
and decreases with increasing $|l|$. For small values of
$|b|$, and still for HM stars, there is also a dependence on $l$, the velocity
dispersion being higher at the centre and decreasing with increasing
$|l|$, as expected. This decrease is stronger for low than for high
$|b|$. These trends can also be found for LM stars, but to a much
less pronounced. The IM
stars have an intermediate behaviour, as expected. 

We thus reproduce well the main observational characteristics
  found by N13b. In the lowest metallicity bin, however, the 
$\sigma$ stays considerably flatter with $l$ than in N13b. This
discrepancy is even stronger when we compare with the Z16 data. 
Indeed, the Z16 data give results qualitatively in agreement with
\citeauthor{Ness.P.13b}, but quantitatively showing a smaller
difference between the kinematics of the various
populations. This could well be understood by the different selection
criteria of the N13b and Z16 samples (inclusion or not of
foreground stars, choice of metallicity bins, latitudes,
etc.). Nevertheless, the effect of the various selection criteria on the
data is well beyond the scope of this letter and
will be left for future work. However, this difference between the two
observational data sets helped us focus on the initial [Fe/H] used our
simulations (Sect.~\ref{subsec:code}) and  
we made a new calculation, starting the chemical
evolution from zero metals initially. This 
produced a considerably stronger decrease of $\sigma$ with $|l|$,
although always less than the corresponding decrease in the higher
metallicity bins. We can 
thus achieve a good qualitative agreement. It is difficult, however, to
obtain a quantitative agreement and to find,
even approximately, what the best initial metallicity value is.
We thus again limit ourselves to a qualitative agreement.  

We conclude that there is very good qualitative agreement
between our simulations and observations, and
we stress that this is found directly
from the simulations, without having to add by hand any further stars,
implying that our scenario of the galaxy formation history 
gives a right mix of the various types of stars.

\subsection{Stellar ages and star formation (SF)}
\label{subsec:age-jj-plot}
 
In the middle panel of Fig.~\ref{fig:various} we plot the mean
star formation time after binning stars by metallicity and longitude.  
We see clearly
that the stars in the low metallicity bin were born preferentially
between, on average, one or two Gyr
from the beginning of the simulation, i.e. around $t_{bm}$
(1.4 Gyr),
while those in the HM bin were 
born considerably later, between, on average, 4 and 7 Gyr ago. Note
also that there is not 
much dependence of these birth times on $|b|$.

The right panel of Fig.~\ref{fig:various} shows the star formation
history of stars within the inner 3.5 kpc (SFH$_{3.5}$), i.e. the
histogram of the stellar mass born at a given time. This is done
separately for our three metallicity bins, i.e. LM, IM, HM 
and the yet higher
metallicity HHM bin. 
We note that the LM and HM have SFH time distributions of quite
different forms. The former were born much before the latter, in good
agreement with what was shown in the middle panel of
Fig.~\ref{fig:various}. Furthermore, 
the birth times of the LM stars are quite concentrated around 
$t_{bm}$ (1.4 Gyr), suggesting that these star formations are due to a  
starburst, presumably from strong inflow of gas 
during the merging. On the other hand, those of the
HM stars are very spread out, starting from 7 Gyr ago and continuing up to the
end of the simulation. 
The IM stars are intermediate, as expected. 

\subsection{Spatial distribution of metallicity-defined populations}
\label{subsec:spatial-distrib}
 
\begin{figure}
  \centering
  \includegraphics[scale=0.52]{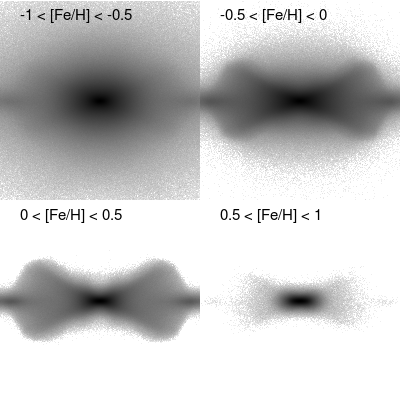}
  \caption{Edge-on views of the baryonic projected surface density in four
    metallicity bins. 
    The side of each of the four 
    squares covers $\pm$3 kpc from the centre of the galaxy.  
    }
  \label{fig:sideon4}
\end{figure}

\begin{figure}
  \centering
  \includegraphics[scale=0.52]{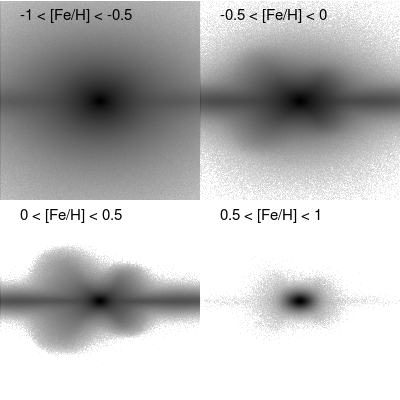}
  \caption{($l$,$b$) perspectives of the baryonic projected surface density in four
    metallicity bins. The bar major axis is at 27$^{\circ}$
    from the line of sight to the centre of the galaxy, as in the MW
    and positive $l$ is on the left. 
    }
  \label{fig:lb4}
\end{figure}

To understand the spatial distribution of populations with different
metallicities we viewed them 
side-on, i.e. edge-on with
  the line of sight in the equatorial plane and along the bar minor
  axis (Fig.~\ref{fig:sideon4}). In this figure we wish to view 
the bar and its B/P structure in an optimal manner and not to compare
with observations.  
We thus exclude stars from the foreground and
background disc component, using a cut-off of
$|\Delta y|$$<$0.5  kpc, where $y$ is the distance of the `star' from
the plane perpendicular to the line of sight and through the centre of
the galaxy. 
On the contrary, in Fig.~\ref{fig:lb4} we view the bar in $l$, $b$
coordinates, i.e. as viewed from the Sun, and introduce only the 
cut-off of $R$$<$3.5 kpc. 
We then split, both for Figs.~\ref{fig:sideon4} and \ref{fig:lb4} all
stars in four metallicity bins -- LM, IM, HM and HHM. 

The difference between the distributions corresponding to different
metallicity ranges is striking. The
lowest metallicities are distributed in a spheroidal-like shape
flattened in the vertical direction, and including a low density disc,
which has a thick and smooth outline in the $z$ direction. They do not
show any, or very little, X-shaped structure, in good agreement with
observations \cite[e.g.][]{Ness.P.12, Uttenthaller.SNRLC.15, 
Rojas-Arriagada.P.14, Kunder.P.16}.    

The IM stars 
show a clear X-shape
embedded in a spheroidal-like or boxy-like part, implying that the
stars in this bin come from two quite different populations and
components. The HM bin 
has only the X-shape and a clear underlying disc component. These two
panels show that the 
X-shape is of the off-centred type (OX, \citealt{Bureau.AADBF.06}),
i.e. that the
branches of the X join only in pairs, two on either side of the centre
($>$----$<$), contrary to the centred X (CX), where all four
branches join in the centre ($>$$<$). This is the case for about half
of the \citeauthor{Bureau.AADBF.06} sample 
of X-shaped bulges.   

Finally, the distribution of the material in the last metallicity bin
(HHM) is centrally concentrated and 
its vertical extent is small compared to its horizontal extent. Although this
definitely needs further study, it is tempting to associate it with a
discy pseudo-bulge (see e.g. \citealt{Athanassoula.16} and
\citeyear{Athanassoula.08} for definitions and reviews).      

These density distributions show that the bar/bulge region
is complex, with many co-existing stellar populations, while the ratio of
their mass is a function of location. This also is in good agreement with
observations \cite[e.g.][]{Ness.P.13a}. 

\section{Summary and discussion}
\label{sec:summary}

We use numerical simulations that follow the formation of a disc galaxy
subsequent to a major merger of two protogalaxies with extended gaseous
haloes. In a previous paper (A16) we showed that such events could
form realistic disc galaxies. We thus apply such a model to the central
bar/bulge region of our Galaxy.

Metallicity distribution functions \citep{Ness.P.13a}
imply the
existence of more than one stellar population in this region. We thus
analysed, as in observations, separately three populations,
with metallicities -1.$<$[Fe/H]$<$-0.5 (LM),
-0.5$<$[Fe/H]$<$0 (IM) and 0$<$[Fe/H]$<$0.5 (HM), respectively.     

As in observations, we limited ourselves to the inner 3.5 kpc so
as to concentrate on the bar/bulge region. We compared the velocity dispersions of the LM and HM populations (IM
stars have an intermediate behaviour) and found them quite different.
This implies that they have quite different kinematics. LM
stars have large velocity dispersions, which, furthermore,
show little, or no, dependence on either latitude or longitude.
HM stars located near
the equatorial plane have
roughly the same velocity dispersion as the LM ones, but this dispersion
decreases substantially with increasing $|l|$, or b. The
decrease with $b$ is stronger for smaller longitudes.
  
We thus find excellent qualitative agreement 
between our simulations results and those of spectroscopic observations, 
concerning the kinematics, the metallicities and their link. This argues
that the properties we are discussing must be generic to disc galaxies
with boxy/peanut/X bulges, because our model, although it gives a
reasonable fit to a number of MW quantities and properties, was
not specifically built for this purpose. 

We also followed the ages and the SFH$_{3.5}$ separately for each of our 
metallicity defined populations. We find that the LM stars were born
early on and in a relatively narrow time range centred roughly on
$t_{bm}$ (1.4 Gyr).
On the other hand, the HM stars are born later, up to considerably later,
and over a much broader time range which extends to the end of the
simulation. 

Finally we inspected the morphology of each of these populations
separately. We found that the LM population makes a flattened
spheroidal-like object, which could be considered as a compound of the
thick disc, a stellar halo and (whenever present) a small classical bulge.  
The IM and HM populations show clearly an X-shape when viewed 
side-on and are the two only populations to do so.

A clear conclusion from all our results is that the central
region of our Galaxy, and therefore of a number of external barred
galaxies, is very complex, which can be understood 
by realising that a number of components  
are cohabiting in this area. These components have widely different
morphological, kinematical and chemical properties and the properties
of any region depend on the relative density of these components in
it. Such a mix may not be obvious to disentangle quantitatively, but is a natural
consequence of the dynamical processes governing the formation of the
MW.

In A16 we had decomposed the baryons as a function of their birth time
in five different populations. Using morphology, radial projected
density profiles and circularity parameters, we found that stars born
in the individual protogalaxies, i.e. before the merging, undergo
violent relaxation during the merger and end up mainly in a triaxial classical
bulge and a stellar halo. Stars born during the merging period contribute partly to a
similar though more extended spheroid and partly to a thick disc and
the bar that forms from it. 
Stars born in the beginning of this period mainly
contribute to the classical bulge and only little to the thick disc. 
On the other hand, when we consider stars born at times nearer to the
end of the merging period, 
it is the thick disc that is the main contributor. At times after the
end of the merging period (more precisely for times after $t_{bd}$)
the gas accreted from the gaseous halo forms a thin disc and its
substructures such as spirals, or a bar, or a boxy/peanut bulge. This
picture is of course a very simplified description of how the stellar
populations form, because for example stars born after $t_{bd}$ (2.2
Gyr) in the
thin disc could become members of the thick disc after being perturbed by
small dark or luminous satellites, or by spirals. Furthermore, material
from the gaseous halo will fall in at all times, and not only after
$t_{bd}$. Thus the boundaries set by the landmark times are not sharp 
and the whole evolution can be seen as a continuous sequence from an all-spheroid to
an all thin disc formation, with strong changes at the landmark
times. 

The results we find here are well compatible with this evolution
picture proposed in A16. We find that the LM stars are old, and 
are born in a relative short time range around the
merging time. They have larger velocity dispersions than HM stars,
compatible with them being stars partly in a spheroid and partly in a
thick disc. Most important,
their 2D projected density distribution has the right shape for such
components. 

The HM stars were born on average much
later than the LM ones, and over a much more extended range of times,
extending all the way to the present time. They generally have smaller
velocity dispersions, as would be expected for disc stars. For
$b$=0$^{\circ}$, and in general up to roughly $|b|$=4$^{\circ}$,
their velocity dispersion decreases with increasing $|l|$ (left panel
of Fig.~\ref{fig:various}, as expected
for a disc. Most important, the corresponding 2D projected density
distribution is that of a disc population, some of it having undergone
the bar instability and then formed a clear X-shape. 

Thus in the bar/bulge region we find a classical bulge component, a
thin and a thick disc component, together with the corresponding 
bar and boxy/peanut/X bulge, and, most probably, a discy bulge,
although to affirm the latter we need yet higher resolution simulations. 
The ratio of 
masses of these components vary from one barred galaxy to another, so
the one we have here may not correspond accurately to that of the MW.   

It is important to stress that none of the components is strictly
confined to a given bracket, be it in time, or age, or metallicity. Thus
at any location there is mix of populations. This cohabitation
explains the links of metallicity with kinematics (as 
found in observations) and also with
morphology, as stressed here. Regions that are dominated by the
spheroid population will have both spheroid kinematics and spheroid
metallicities, while regions which are dominated by disc population
will have both disc kinematics and disc metallicities. This makes the link
between kinematics and metallicities obvious, since we can not have e.g. spheroid
kinematics and disc metallicities in the same component. Thus such
links should be found 
also in external galaxies having a B/P structure viewed edge-on.


\section*{Acknowledgements}

We thank A. Bosma, C. Chiappini, K. Freeman, M. Ness and G. Zasowski
for stimulating and useful discussions and the referee for a
constructive report.
We acknowledge financial support from the CNES
(Centre National d'Etudes Spatiales - France), as well as 
HPC resources from GENCI - TGCC/CINES x2016047665 (program SINGSONG) 
and from the Mesocentre of Aix-Marseille Universit\'e
(program  DIFOMER).

\bibliographystyle{mn2e.bst}
\bibliography{mdf_chemi_all.bbl}

\label{lastpage}

\end{document}